\documentstyle[psfig,raulbib,times]{mn}

 \def\la{\hbox{
\raise.35ex\rlap{$<$}\lower.6ex\hbox{$\sim$}\ }} \def\ga{\hbox{
\raise.35ex\rlap{$>$}\lower.6ex\hbox{$\sim$}\ }}

     \def\W2{{\cal W}}

\begin{document}

\title[Galactosynthesis Predictions at High Redshift]{Galactosynthesis
Predictions at High Redshift} \author[Buchalter, Jimenez \&
Kamionkowski] {Ari Buchalter$^1$, Raul Jimenez$^2$ \& Marc
Kamionkowski$^1$ \\ $^1$California Institute of Technology, Mail Code
130-33, Pasadena, CA 91125.\\ $^2$Department of Physics and
Astronomy, Rutgers University, 136 Frelinghuysen Road,
Piscataway, NJ 08854--8019 USA.}

\maketitle

\begin{abstract}
We predict the Tully-Fisher (TF) and
surface-brightness--magnitude relation for disk galaxies at
$z=3$ and discuss the origin of these scaling relations and
their scatter.  We show that the variation of the TF relation
with redshift can be a potentially powerful discriminator of
galaxy-formation models.  In particular, the TF relation at
high redshift might be used to break parameter
degeneracies among galactosynthesis models at $z=0$, as well as to
constrain the redshift distribution of collapsing dark-matter halos,
the star-formation history and baryon fraction in the disk, and the
distribution of halo spins.
\end{abstract}


\section{Introduction}

With the advent of numerous 10m-class telescopes, adaptive
optics, and high-$z$ galaxy surveys, it is becoming
possible to learn about the detailed properties of galaxies at
high $z$.  In the future, we may expect improved IR
sensitivity with NGST and 30m or even 100m telescopes, and this
should usher in an era of high-precision kinematic and
photometric studies of high-$z$ galaxies.
Disk galaxies are particularly
important to our understanding of galaxy formation and
evolution, as they are believed to undergo a relatively smooth
formation process, and possibly serve as building blocks in
the formation of other galactic systems through mergers.

Many authors have investigated galactosynthesis models
(e.g., \pcite{DSS97,MMW97,JPMH98,sp99,bosch00,af00,fa00,ns00}) in which
the properties of disk galaxies are determined primarily by the
mass, size, and angular momenta of the halos in which they form,
and which may contain the effects of supernova feedback,
adiabatic disk contraction, cooling, merging, and a variety of
star-formation (SF) recipes.  In an earlier paper, we
(\pcite{BJK2000}; hereafter BJK), investigated a variety of
galactosynthesis models with realistic
stellar populations and made multi-wavelength predictions
for the Tully-Fisher (TF) relation.  With reasonable
values for various cosmological parameters, spin ($\lambda$)
distributions,
formation-redshift ($z_f$) distributions, and no supernova feedback,
we could produce an excellent fit to the local TF
relation at all investigated wavelengths ($B$, $R$, $I$, and
$K$), as well as to $B$-band TF data at $z=1$, and to
the surface-brightness--magnitude ($\mu$-$M$) relation locally
and at $z=0.4$.  These successes suggest that our simplified
approach captures the essential phenomenology, even if it leaves
out some of the details of more sophisticated models.

In this paper, we investigate the high-$z$ TF predictions
for our model.  We examine the key factors impacting the TF
relation at high $z$, and demonstrate that degeneracies among
parameters in galactosynthesis models at $z=0$ can be disentangled
at high $z$.  Specifically, information about the $z_f$ and
$\lambda$ distributions, as well as the SF history and baryon
fraction in the disk, can be gleaned from the high-$z$ TF relation.

\section{The Models}

Here, we briefly review the main ingredients of our model and refer
the interested reader to BJK for a
comprehensive description.  We use the spherical-collapse
model for halos, the distribution of halo-formation times
{}from the merger-tree formalism \cite{lc93,lc94}, and a joint
distribution in $\lambda$ and $\nu$, the peak height
\cite{HP88,ct96}.  Halos are treated as isothermal
spheres with a fixed baryon fraction and specific angular momentum,
and their embedded gaseous disks, assumed to form at virialization,
have an exponential density profile.  We implicitly assume that
the halos of spiral galaxies form smoothly, rather than from
major mergers (BJK; \scite{EL96}).  An empirical Schmidt law relates
the star-formation rate (SFR) to the disk surface density \cite{K98}.
A Salpeter initial mass function, a prescription for chemical
evolution, and a synthetic stellar-population code \cite{JPMH98}
provide the photometric properties of disks at any $z$.

The model is defined by cosmological parameters and by the
time when the most massive progenitor contains a
fraction $f$ of the present-day mass, when a halo is defined
to form.  We found that excellent agreement with current data
was obtained by our 'Model A', a COBE-normalized $\Lambda$CDM
cosmogony ($\Omega_0=0.3$, $h=0.65$, $\Omega_b h^2 =0.019$) with
$f=0.5$.
The TF relation in this model relied on both halo properties and
SF in the disk. The local TF scatter arose
primarily from the $z_f$ distribution, and
secondarily from chemical evolution and the $\nu$-$\lambda$
anticorrelation.  In this model, disk formation
occurs primarily at $0.5 \la z \la 2$ and
the TF slope steepens and the zero points get fainter from $z=0$ to
$z=1$.  Moreover, the amount of gas expelled from or
poured into a disk galaxy in this model is relatively small and
the disk and halo specific angular momenta are equal.

A suite of other models that give good fits to the
observations at low $z$, can also be found.
To illustrate, we examine two alternatives that yield
comparably good fits to the slope, zero-point, {\em and} scatter
of the TF relation at $z=0$ in $B$, $R$, $I$, and $K$. The first
is a CDM model with $\Omega_0=1$,
$h=0.65$, constant metallicity given by the solar value,
$\lambda=0.05$ for all disks, a power-spectrum
amplitude $\sigma_8=0.5$, an empirical shape parameter
$\Gamma=0.2$, and $f=0.5$.  High-$\Omega_0$ models generally
produce disk galaxies too faint to lie on the local TF relation.
To compensate for this, we assume $\Omega_b h^2 = 0.045$, and
thus term this the 'high-$\Omega_b$' model.

Our second alternative is a $\Lambda$CDM cosmogony like Model A,
but with metallicity held constant at the solar value,
$\lambda=0.05$ for all disks, and $f=0.9$.  This
results in a narrow distribution of formation times peaked
around $z\sim 0.2$ (for $L_*$-type disks), with
appreciable ongoing formation today and almost no halos forming
earlier than $z=1$ (cf., Figure 1 of BJK).  We thus denote
this the 'low-$z_f$' model.  This model
will produce extremely young and bright disks (cf.,
Figure 5 of BJK)\footnote{We note that bulges, however, may be
present at high $z$.}. To compensate for this, we reduce the
efficiency of SF in our Schmidt-law prescription by 50\%
[see equations (2)--(4) in BJK].  This lower SF
efficiency is plausible given a possible inconsistency between
the Model-A predictions for the absolute disk gas
fractions and those observed (see BJK for details).

\begin{figure*}
\centerline{ \psfig{figure=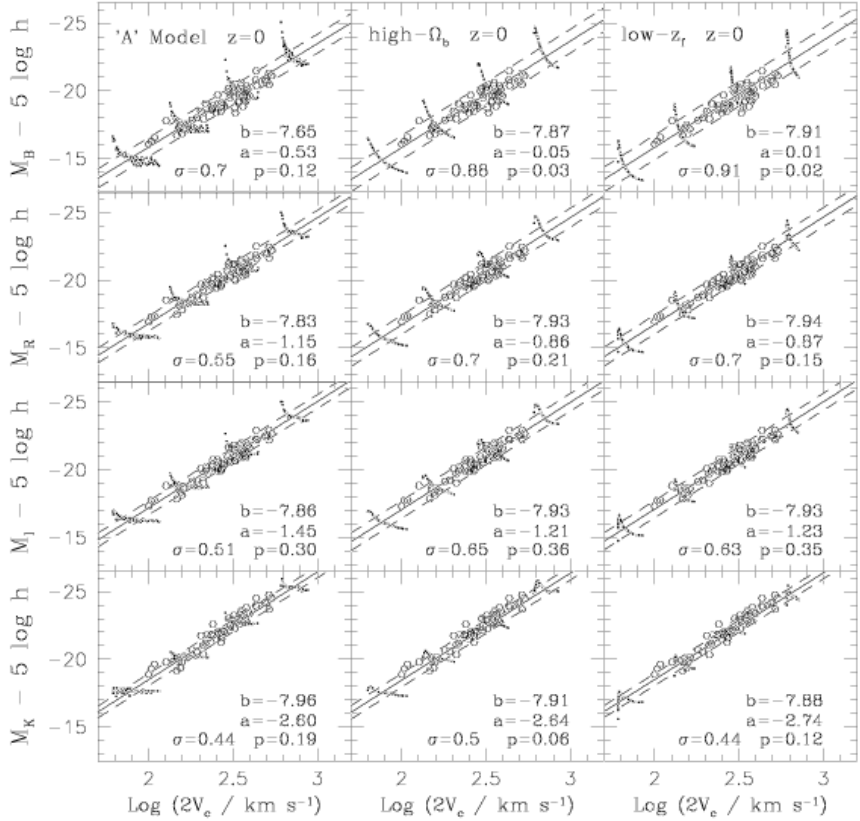,height=12cm,angle=0}}
\caption{Predicted $B$-, $R$-, $I$-, and $K$-band TF relations at
$z=0$ for the A model, high-$\Omega_b$ model, and low-$z_f$ model.
The scattered dots show the results for fixed masses
of $10^{10}$, $10^{11}$, $ 10^{12}$, and $10^{13}$ ${\cal M}_\odot$.  The
solid lines are the best-fitting TF relations, with zero points and
slopes given by $a$ and $b$, respectively, while the dashed lines show
the 1$\sigma$ scatter, denoted in each plot by $\sigma$.}
\label{fig1}
\end{figure*}

Figure \ref{fig1} depicts the $z=0$ predictions for the A,
high-$\Omega_b$, and low-$z_f$ models, in the $B$, $R$, $I$, and
$K$ bands.  The open symbols represent extinction-corrected data
{}from \scite{TPHSVW98}.  In each plot, the data are fit to the
corresponding model, and $p$ gives the probability for the data
given the model. Since the data
have excluded spirals that show evidence of merger activity or disruption,
we exclude from our predictions those galaxies with $B-R < 0.3$.

Each of the three models yields a reasonable fit
to the slope and normalization of the TF relation in all wavebands.
Moreover, the predicted scatter in the $B$, $R$, $I$, and $K$
bands roughly agrees with the
observed values of 0.50, 0.41, 0.40, and 0.41, respectively, with the
largest discrepancies occurring in the bluer bands, for reasons discussed
in BJK.

\section{The TF Relation at High $Z$}
\subsection{TF Scatter for a Fixed Mass}

\begin{figure*}
\centerline{ \psfig{figure=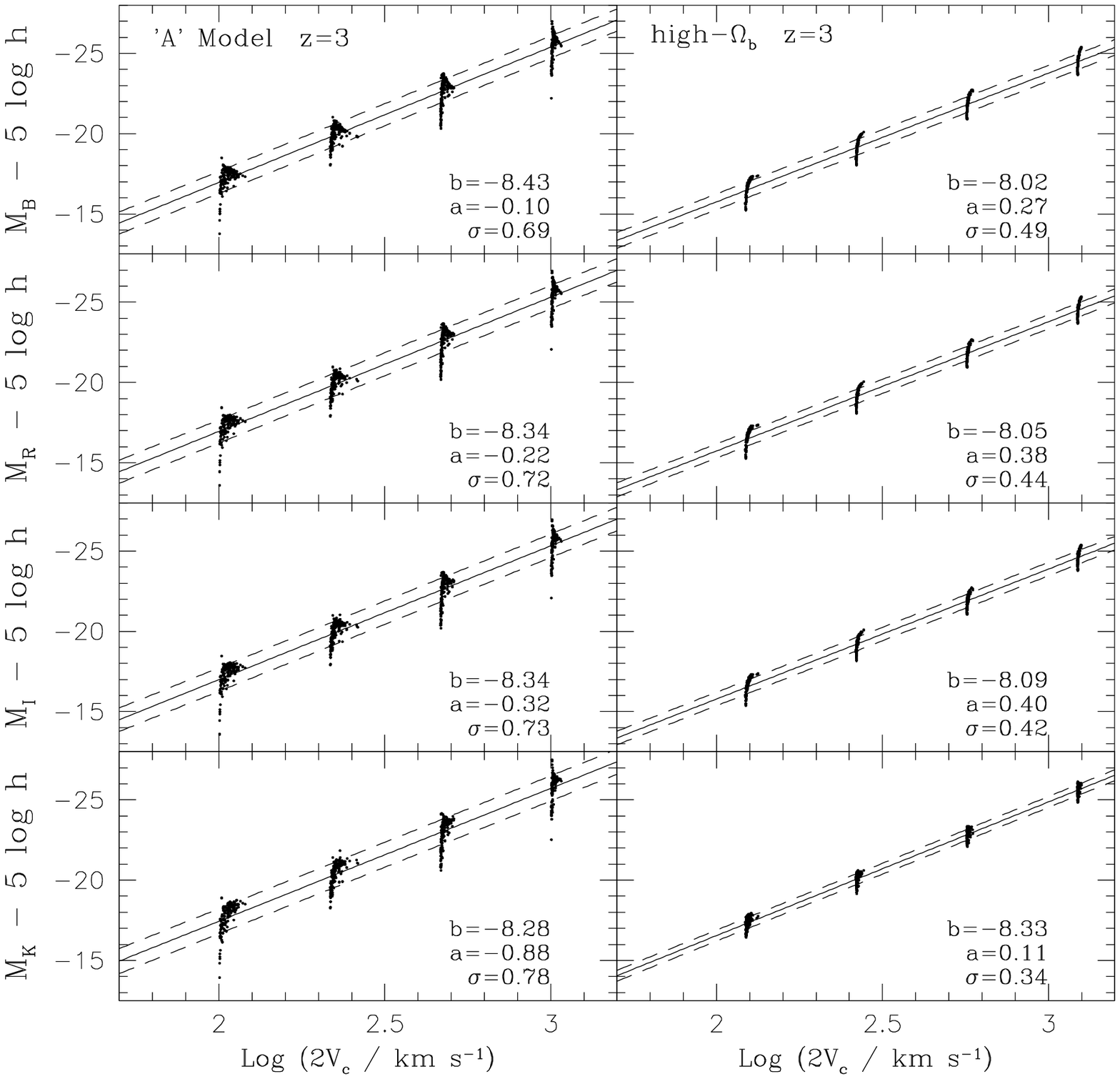,height=12cm,angle=0}}
\caption{TF predictions of the A model and high-$\Omega_b$ model at $z=3$. The
low-$z_f$ model predicts essentially no disks to have formed at $z=3$ at these
scales.
}
\label{fig2}
\end{figure*}

Figure \ref{fig2} shows predictions for our A and high-$\Omega_b$
models at $z=3$. At this $z$, the low-$z_f$ model fails to
form disks.
In the left-hand panels, the points for a given
mass scatter in a nearly vertical direction in contrast to the more
horizontal scatter in the $z=0$ results (see BJK Figure
13), particularly in redder bands.  For $z=0$, the TF
scatter arises primarily from the spread in $z_f$, with
the earliest-forming objects of a fixed mass having higher circular
velocities.
This is because for an evolved galaxy that has
converted most of its gas to stars, the light from giants serves as a
faithful tracer of the total mass, regardless of the precise age of
the galaxy.
As one goes either to higher masses in a given waveband, or to bluer
bands at a fixed mass, the scatter in the local TF relation for a
fixed mass does begin to curve upward (i.e., later-forming disks
appear brighter) to form a 'magnitude peak,' since in both cases one
becomes increasingly biased towards younger, brighter systems.

At $z=3$, the scatter still owes mainly to the spread in $z_f$, and
Figure \ref{fig2} still indicates that at $z=3$, the very
earliest-forming objects do possess somewhat higher circular
velocities for a given mass.
(Note in the left panels the 'peak' of points formed at the high-$V_c$
end of the four scatter plots for the different masses.)  The galaxies
that populate the TF relation at $z \ga 3$, however, are the small
minority that constitute the exponentially decaying high-$z$
tail of the adopted $z_f$ distribution (BJK Figure
1). Thus, the formation of these halos spans a very short epoch,
so that halos of a given mass will have roughly the same $V_c$,
resulting in the tight horizontal scatter for each mass.  The large
vertical scatter arises from the fact that, though these galaxies will
eventually form the lower envelope of the present-day TF relation,
comprising the oldest and (in bluer bands) faintest objects of a given
mass, at $z=3$ even the least massive of these galaxies are typically
no older than a few hundred million years, and the more massive
systems younger still.
Thus, systems on the 'young' side of the $M$ peak have
formed so close to the epoch of observation that they simply have not
had sufficient time to build up appreciable stellar populations; given
their high gas fractions and SFRs, even a small
increase in their age results in a substantial increase in their
stellar luminosity.

At high $z$, disks with circular velocities
200--500 km s$^{-1}$ (masses $\sim 10^{13}$ ${\cal M}_\odot$)
correspond to rare fluctuations. Still, at these $z$, one
does expect to find
several such objects within the horizon in a $\Lambda$CDM cosmogony,
and since these will be the brightest objects in the sample, we
include them in the discussion here.  We also reiterate that the disks
that form at $z \ga 3$ are not the dominant population of disks at
$z=0$, as most disks at $z=0$ probably formed
at $0.5 \la z \la 2$ (see, e.g., BJK). Disks forming at higher $z$
will delineate the lower envelope of the TF relation at $z=0$; for a given
mass, these
older, higher-$V_c$ systems will be fainter since they will be the
first to run out of fuel.

\subsubsection{Overall TF Relation}

The trend of steeper TF slopes and fainter zero-points in going
{}from $z=0$ to $z=1$ is seen to continue here out to $z=3$.
Specifically, in the $B$ band, the $z=3$ TF relation goes from $\sim
2$ mag brighter than its local counterpart at $\log{2 V_c}=3$ to less
than 1 mag brighter at $\log{2 V_c}=1.7$. In $K$, the high-$z$
prediction is about 1 mag fainter at $\log{2 V_c}=1.7$ and comparable to
the local prediction at $\log{2 V_c}=3.0$.  This is because
for fixed $V_c$, halos at high $z$ will be younger, smaller, and
have lower $\lambda$ than halos today.  In massive systems most of
the SF activity takes place at
early times. Thus, younger systems will be significantly brighter,
particularly in bluer bands, even though much of their gas remains
unprocessed.  Smaller, low-$V_c$ systems, on the other hand, do not
convert gas to stars as rapidly and are thus fainter than their local
counterparts.

The {\em overall} A-model $B$-band scatter in the high-$z$ TF
relation is comparable to that at $z=0$.
In $K$, however, the scatter at $z=3$
has increased by a factor of two. Moreover, the variation
in the TF scatter among different wavebands is not as great at high
$z$ as was seen in the local case.  This is because (a)
disks at $z>3$ are likely to be in a
starburst-dominated phase since there is very little cosmological time
for even single stellar populations to decay; (b)  at such early
times the spread in mass of gas turned into stars is very large for
different $z_f$ and $\lambda$ [see eq. (7) of
\scite{HJ99}]. Therefore, it seems natural that all bands reflect the
maximum possible spread. By contrast, there is little change in TF
scatter from $z=0$ to $z=1$, as the
majority of galaxies are sufficiently evolved by $z=1$ (Figures
13 and 14 in BJK).
Furthermore, the scatter
obtained here should be considered as an upper limit since, in
practice, $M$-limited surveys may not detect arbitrarily young,
and therefore faint, objects, particularly at lower
masses.\footnote{Our models employ a minimum time step of $10^7$
yr---corresponding to the timescale for giant-molecular cloud
destruction (e.g., \pcite{jim00})---that
effectively dictates the faintest objects at a given mass,
and therefore the maximum possible scatter.}  However,
different realizations of a given model can generally produce
results that vary by $\la 0.1$ mag in the TF scatter.  Given
the simplifying assumptions of the model, attention should be
paid to the overall magnitudes of and relative differences
between the slopes, normalizations, and scatters, rather than
their precise values.

\subsubsection{Spin--Peak-Height ($\nu$-$\lambda$) Anticorrelation}
In BJK, we also noted the slight
reduction in TF scatter at $z=0$ in Model A arising from the
existence of a weak
anticorrelation of $\lambda$ with $\nu$, and thus with $z_f$.
This reduction in TF scatter was only about 0.15
mag in $B$ and $0.05$ mag in $K$ for Model A, and lesser or no
effect was seen in some other possibly viable models.  To investigate
the effect of the joint $\lambda$-$\nu$ distribution,
we re-ran the models in Figure \ref{fig2} with
$\lambda=0.05$ for all halos.  This produces TF predictions
with shallower slopes and brighter
zero-points. This is due to the fact that the
$\lambda=0.05$ value is higher (lower) than the full model would
typically assign to high-mass (lower-mass) galaxies, which thus become
correspondingly fainter (brighter) in the fixed-$\lambda$ model. More
interestingly, using a fixed $\lambda$ at high $z$ {\em reduced}
the overall TF scatter by about 0.2 mag in all bands at $z=3$, as
compared with using the joint distribution in $\lambda$ and $z_f$.
This is in direct contrast to the $z=0$ case, where a fixed $\lambda$
produced a comparable or larger scatter, as described above. This is
due to the vertical nature of the scatter in the TF for a fixed mass
at high $z$. In this case, higher-$\lambda$ systems are still fainter,
but their lower $z_f$ (already confined to only a narrow
range) can only further scatter them to lower luminosities, not to
lower $V_c$ as is in the local case, thereby increasing the
scatter.
This suggests that the scatter in the TF relation at high
$z$ could be used to gauge the strength of the actual
$\lambda$-$z_f$ anticorrelation.

\subsection{The Surface-Brightness--Magnitude Relation}

\begin{figure}
\centerline{ \psfig{figure=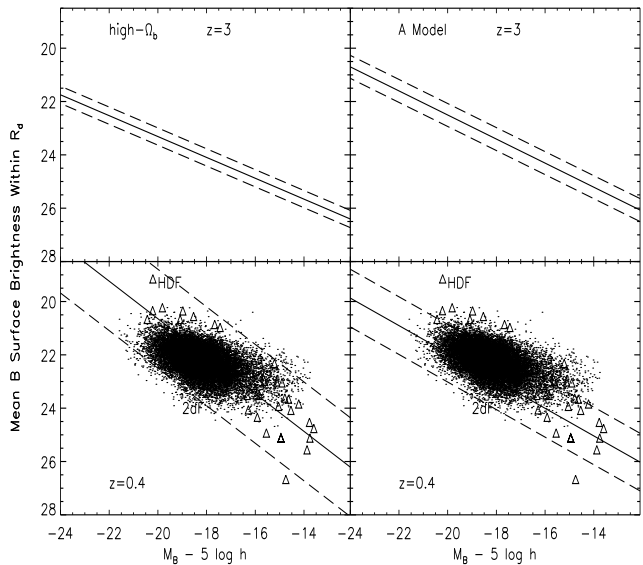,height=8cm,width=9cm,angle=0}}
\caption{Predicted surface-brightness--magnitude relation for disk
galaxies at $z=0.4$ and $z=3$ for the A model and high-$\Omega_b$ model.
The solid lines are the best-fitting
results with a 1$\sigma$ scatter given by the dashed lines. Note that
our model does not include a bulge component.}
\label{sb}
\end{figure}

Figure~\ref{sb} shows the surface-brightness--magnitude relation
at $z=0.4$ and $z=3$ for the A and high-$\Omega_b$ models,
The dots are from 2dF \cite{drivercross00}, and the
open triangles from the HDF \cite{driver99}; the latter have $z\simeq
0.4$.  Since our model does not include a
bulge component, the predicted $\mu$ may
underestimate observed values if the disk and bulge components
cannot be resolved.

For Model A, the slope of the relation at $z=3$ is slightly
shallower than the local
relation, with $\mu$
nearly identical to the low-$z$ case for $M_{B} - 5 \log h > -16$
and fainter by about 1 mag for $M_{B}- 5 \log h = - 24$.  Thus, at
these scales, the trend towards higher luminosities and smaller disk
scale lengths is offset by cosmological $\mu$ dimming.
A more drastic contrast is seen in the high-$\Omega_b$ model,
where the predicted $\mu$ at the bright end varies
by almost 4 magnitudes. This owes largely to substantial brightening
that occurs at late times as the high baryon fraction is converted into
starlight.

For both models, we find that the scatter at $z\geq 3$
is much tighter than for the local
case because the spread in the radii of disks at $z>3$ is
smaller than at low $z$. For example, at $z\sim 0$, disk sizes span
a range roughly between 1 and 100 kpc, while at $z\sim 3$ they range only
{}from about 0.5 to 15 kpc in size. On the other hand, the
respective spreads in
$M$ differ by no more than a factor of two.
The small spread in the relation then reflects the small spread in
disk sizes predicted in our adopted structure-formation scenario.
Thus, it will be interesting to learn whether this
narrow spread is in fact observed at high $z$. If a much wider spread
is observed, this might indicate that either disks at $z>3$ are more
substantially affected by merger activity or that they are simply
larger; i.e., the spread in sizes predicted by our simple isothermal
spherical-collapse model with angular-momentum conservation is simply
too narrow.  Note that the $\mu$ predictions in general
are more sensitive to assumptions regarding $\lambda$ than are TF
predictions, thus providing a strong complementary constraint on this
aspect of the model.

\subsection{Addressing Possible Degeneracies}

We now address how high-$z$ predictions can be used to
distinguish among galactosynthesis models that are degenerate at
$z=0$.
The right panels of Figure \ref{fig2}
show TF predictions for
the high-$\Omega_b$ model. At these mass scales,
the low-$z_f$ model predicts essentially
no disks (and thus no TF relation) to be present at $z=3$, certainly
a drastically different prediction than that of the other models,
despite their close agreement at $z=0$.

The precise slopes and zero-points for the A and high-$\Omega_b$
models differ more appreciably than
in the $z=0$ case, with the latter model yielding predictions
that are significantly fainter at these scales. This owes both
to the different normalizations of the fluctuation spectrum in the two models,
and to the fact that the lower SF efficiency in the
high-$\Omega_b$ model results in fainter disks at early times.
More importantly, the predicted TF scatter in the high-$\Omega_b$ model is
substantially less than that of the A model. This is because
(a) a higher value of $\Omega_0$ implies a narrower age range for disks forming
at $z\geq 3$; and (b) the scatter due to a spread in $\lambda$ is
excluded in this model.

Alternatively, one could envision
a high-$z_f$ model in which disks were somehow brightened at later
times, perhaps through late infall of fresh gas. In this case,
there might be galaxies that are reasonably evolved, even
at these high $z$, so that many galaxies would be found on the old
side of the $M$ peak. Thus, the TF scatter for a fixed mass for such
objects would
more closely resemble that of Figure \ref{fig1}, rather than the more
vertically-oriented scatter of Figure \ref{fig2}, and presumably also
lead to a smaller overall TF scatter.
The differences among the high-$z$ predictions explored
here for the TF relation and its scatter amount to about 1--2
mag, and 0.2--0.4 mag, respectively, so these models should
be distinguishable with
high significance given data of the type in Figure \ref{fig1}.
The model predictions for the $\mu$-$M$ relation
can also be used to address the degeneracy. In Figure~\ref{sb}, we see that,
owing to the large scatter, the two models are difficult to distinguish
on the basis of low-$z$ predictions, but the tightness of the predicted
relation at higher $z$, if in fact observed, would make
this test a much more powerful discriminator.

\section{Discussion}

We have presented high-$z$ predictions for the TF relation,
which may help to discriminate among different scenarios for
galaxy formation. The evolution of the spread in the TF
relation can probe the spread in halo $z_f$, as well as
SF processes in the disk.
Specifically, for observations of evolved systems
at redder wavelengths, the scatter essentially decouples from the
luminosity axis, while observations at high $z$ and/or in bluer
bands are more sensitive to the disk's age,
resulting in a scatter that couples more closely to the
luminosity axis of the TF relation.  The predicted
$\mu$ for a given $M$ are fainter and
tighter at high $z$ than at low $z$.

Our models assume
that galaxies in the high-$z$ universe are those in the
high-$z_f$ tail of the present-day distribution.
The circular speeds we predict are those for the disk, and so
far it has proven difficult to extract these reliably from
general SF population at $z=1$.
Recent data seem to suggest that beyond the local universe, ``normal''
galaxies actively form stars primarily in their small cores (see,
e.g., \pcite{sim98,rix97}).  Linewidth measurements of SF
galaxies at $z=1$ and $z=3$ are found to effectively sample only the
core dispersion, yielding typical values of roughly 80 km s$^{-1}$
with a scatter of about 20--30 km s$^{-1}$ over a range of several
magnitudes. Unless improved measurements can
succeed in sampling the disk at larger radii (or at some other means
of tracing the potential), a proper TF relation may not be easily
defined.

Progress might be made by focusing on the lower, red envelope
of the TF relation, rather than
attempting to fill in the TF plane.
The locus of the oldest and reddest points in the TF plane generally
lie themselves on a well-defined line that serves as the high-$z_f$
envelope to the relation.  Samples of these old, red objects might be
easier to construct than complete TF samples, which include young,
starbursting objects that may have arbitrarily low luminosities
falling off on the younger side of the $M$ peak.
SIRTF and NGST will have the sensitivity to detect some of
the reddest objects in the sky, even out to high $z$.  If so, a
comparison of the red envelope formed by these old objects with
predictions, could suffice to constrain many of the key aspects of
galactosynthesis models without needing to populate the entire TF
plane. Of course, the effects of dust will need to be taken into
account to so that young, dusty systems are distinguished from
old galaxies.

\smallskip

We wish to acknowledge C. C. Steidel and K. A. Adelberger for helpful
discussions.  This work was supported by grants NSF-AST-0096023,
NSF-AST-9900866, NSF-AST-9618537, NASA NAG5-8506, and DoE
DE-FG03-92-ER40701.

\end{document}